\documentclass[conference,final]{IEEEtran}

\usepackage[utf8x]{inputenc}

\usepackage{amsmath}
\usepackage{amsfonts}
\usepackage{amssymb}

\usepackage{tikz}
\usepackage{pgfplots}
\usetikzlibrary{calc,positioning,shapes,arrows}

\IEEEoverridecommandlockouts

\newcommand{\e}[1]{\ensuremath{\text{e}^{#1}}}
\renewcommand{\j}{\ensuremath{\text{j}}}

\newcommand{\diag}[1]{\ensuremath{\text{diag}\left(#1\right)}}
\newcommand{\trace}[1]{\ensuremath{\text{Tr}#1}}
\renewcommand{\vec}[1]{\ensuremath{\text{vec}(#1)}}
\newcommand{\vecd}[1]{\ensuremath{\text{vecd}(#1)}}
\newcommand{\real}[1]{\ensuremath{\text{Re}#1}}

\newcommand{\mb}[1]{\ensuremath{\boldsymbol{#1}}}

\newcommand{\tT}{\text{T}}
\newcommand{\tH}{\text{H}}
\newcommand{\tF}{\text{F}}
\newcommand{\tTrue}{\circ}
\newcommand{\khatri}{\ast}

\DeclareMathOperator*{\argmin}{arg\,min}

\newcommand{\mysuper}[1]{\raisebox{4pt}[\ht\strutbox]{$\scriptstyle #1$}}

\title{Multidimensional Sparse Recovery for \\MIMO Channel Parameter Estimation}
\author{\IEEEauthorblockN{Christian Steffens\,$^1$, Yang Yang\,$^2$, and Marius Pesavento\,$^1$} 
		\IEEEauthorblockA{\,$^1$Communication Systems Group, Technische Universit\"{a}t Darmstadt, Germany\\
						  \,$^2$Intel Deutschland GmbH, Germany\\
		\{steffens, pesavento\}@nt.tu-darmstadt.de, yang1.yang@intel.com }
\thanks{The authors acknowledge the financial support of the Seventh Framework Programme for Research of the European Commission under grant number ADEL-619647 and the EXPRESS project within the DFG priority program CoSIP (DFG-SPP 1798).}
}

\begin{document}

\abovedisplayskip=0.11cm
\belowdisplayskip=0.11cm
\abovedisplayshortskip=0.09cm
\belowdisplayshortskip=0.09cm
\renewcommand{\baselinestretch}{0.88}\normalsize

\maketitle

\begin{abstract}
Multipath propagation is a common phenomenon in wireless communication. Knowledge of propagation path parameters such as complex channel gain, propagation delay or angle-of-arrival provides valuable information on the user position and facilitates channel response estimation. A major challenge in channel parameter estimation lies in its multidimensional nature, which leads to large-scale estimation problems which are difficult to solve. 
Current approaches of sparse recovery for multidimensional parameter estimation aim at simultaneously estimating all channel parameters by solving one large-scale estimation problem. In contrast to that we propose a sparse recovery method which relies on decomposing the multidimensional problem into successive one-dimensional parameter estimation problems, which are much easier to solve and less sensitive to off-grid effects, while providing proper parameter pairing. Our proposed decomposition relies on convex optimization in terms of nuclear norm minimization and we present an efficient implementation in terms of the recently developed STELA algorithm. 
\vspace{.7em}
\end{abstract}



\section{Introduction}
Mobile communication generally underlies multipath propagation, due to reflection, refraction or scattering on surrounding objects. In a static environment, a propagation path can be characterized by a complex gain coefficient, a propagation delay and an angle-of-arrival (AoA) at the receiver. Knowledge of the channel parameters can be exploited in different ways, e.g., to estimate the user locations by exploiting the characteristics of the dominant multipath components \cite{4558782}, or to estimate the channel response using parametric channel models \cite{sparseMultipath}. 

The channel parameter estimation problem can be formulated as a multidimensional (MD) estimation problem \cite{1547647}-\nocite{mdRare}\cite{sparseUnderwater}. In the simplest form the estimation could be performed by means of an MD Fourier transform, leading, however, to poor resolution in the case of multiple propagation paths. Better performance can be achieved by subspace-based methods. In \cite{1547647} a variation of the MUSIC algorithm for joint MD channel parameter estimation has been presented, which is applicable to arbitrary spatio-temporal sampling. However, since the MD-MUSIC is a spectrum-search based estimator, high estimation accuracy requires a fine MD parameter grid which, in turn, leads to large and complex estimation problems, i.e. in terms of the computational complexity associated with MD-peak search. Search-free methods, such as the MD-RARE \cite{mdRare} and MD-ESPRIT \cite{mdEsprit}, rely on specific structures in the spatio-temporal sampling, e.g., uniform or centrosymmetric sampling.
 
For proper performance, subspace-based methods further rely on a good estimate of the signal covariance matrix which either depends on a large number of signal snapshots or techniques like forward-backward averaging and smoothing, where the latter techniques require centrosymmetric and uniform spatio-temporal sampling, respectively. A drawback of smoothing techniques is the fact that the effective sampling aperture is reduced which degrades the resolution capability. 

In recent years, sparse recovery techniques came into the focus of parameter estimation research and, similarly to subspace-based methods, have been shown to provide the superresolution property \cite{doi:10.1137/0523074}. However, in contrast to subspace-based methods, sparse recovery techniques do not rely on the estimation of a signal covariance matrix and thus show good performance even in the case of low number of signal snapshots or irregular sampling. Two major categories of sparse recovery methods are greedy algorithms, e.g., orthogonal matching pursuit, and convex relaxation techniques, e.g., LASSO or basis pursuit, where the latter category usually shows better estimation performance at the cost of higher computational complexity. For the application of parameter estimation, a major drawback of sparse recovery methods is the requirement for a fine discretization of the parameter space to a grid in order to achieve a high resolution and avoid basis mismatch or off-grid effects \cite{BasisMismatch}. 
The fine 
parameter grid leads to large dictionary matrices and in turn to large-scale estimation problems. In case of MD parameter estimation, this aspect becomes even more critical, since the dictionary size increases exponentially with the number of parameter dimensions. 

A sparse recovery method for MD channel parameter estimation was presented in \cite{sparseUnderwater}. The presented method relies on joint discretization of all associated parameter spaces, resulting in a large dictionary matrix. Due to its large size, the corresponding estimation problem can hardly be solved by convex relaxation methods, hence the authors propose a greedy method referred to as Space-Alternating Orthogonal Matching Pursuit (SA-OMP) which mainly addresses the computational complexity involved with MD-peak search and grid refinement. As we will show by numerical experiments, SA-OMP suffers from a large bias in estimation performance, due to its greedy nature. In \cite{sparseMultipath} a similar approach of discretizing the MD parameter space was presented. The objective in \cite{sparseMultipath} is, however, not the estimation of the channel parameters, but the estimation of the channel response by exploiting the channel's sparse representation in the parameter space. 

In this paper we propose a novel convex problem formulation which decomposes the MD channel parameter estimation problem into successive one-dimensional estimation problems based on convex relaxation in form of nuclear norm minimization. Due to space limitations we restrict our presentation to the two-dimensional parameter estimation problem, but the extension to estimation problems of higher dimensions is straightforward. In contrast to the MD parameter grid for simultaneous parameter estimation, as presented in \cite{sparseUnderwater}, our method requires discretization of one parameter space at a time only. From the estimation results found in the investigated parameter space, the parameters in the remaining dimensions are estimated successively. The successive estimation approach leads to smaller subproblems as compared to the simultaneous estimation of all channel parameters, reduces the sensitivity to off-grid effects caused by inaccuracies in joint discretization of all parameter spaces, and 
simplifies 
the 
peak search. For efficient computation of our novel problem formulation we furthermore present implementation in form of the recently proposed STELA method, which has been shown to have superior convergence speed as compared to, e.g., gradient methods and, admits parallel implementation \cite{stela}. 


\section{Signal Model}

\begin{figure}
  \centering
  \includegraphics{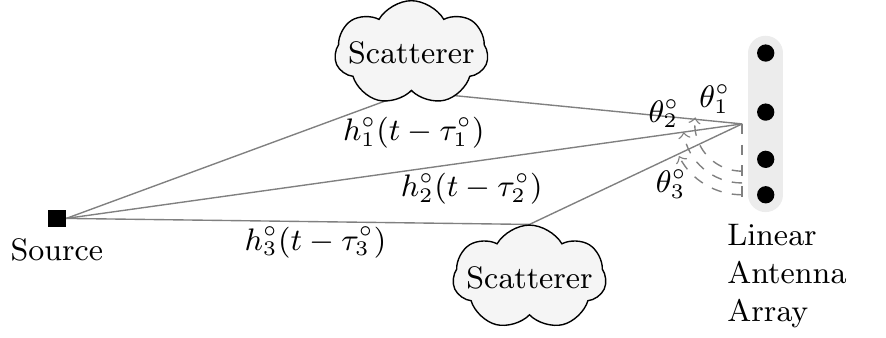}
  \vspace{-0.5cm}
  \caption{Multipath signal model for single antenna transmitter and $M=4$ element antenna array and propagation over $P=3$ paths}
  \label{fig:model}
  \vspace{-0.5cm}
\end{figure}

Consider a single-antenna terminal transmitting a reference signal which propagates over a multipath channel to a receiving linear antenna array, as illustrated in Fig.~\ref{fig:model}. The receiving linear antenna array consists of $M$ omnidirectional antennas with positions given by $r_m \in \mathbb{R}$, for $m=1,\ldots,M$, relative to the first antenna, i.e., $r_1=0$. The reference signal is known to both, the transmitter and receiver. We assume that Orthogonal Frequency Division Multiplexing (OFDM) symbols of length $N$ are transmitted, preceded by a cyclic prefix of duration $T_{\text{cp}}$, sampled at an interval $T_s$, and further assume that the narrowband assumption holds. The OFDM reference signal is denoted as $\mb{x}(l) = [x_1(l), \ldots, x_N(l)]^\tT$ in the frequency domain, where $x_n(l)$ denotes the data symbol, e.g. quadrature phase-shift keying (QPSK) with $|x_n(l)|=1$, on the $n$th subcarrier in the $l$th OFDM symbol. 

The multipath propagation is modeled by $P$ propagation paths. The scattering objects as well as the transmitter are assumed to be located in the farfield region of the antenna array. Each propagation path
is characterized by a complex channel gain coefficient $h_{p}^{\tTrue}$, a propagation delay $\tau^{\tTrue}_{p}$ and an AoA $\theta^{\tTrue}_{p}$, with $p=1,\ldots,P$. The maximum propagation delay is assumed to be smaller than the cyclic prefix duration: $T_{\text{cp}} > \max_{p} \tau^{\tTrue}_{p}$. 

Under the given assumptions, the signal received by the antenna array in time instant $l$ is modeled by the $M \times N$ receive signal matrix 
\begin{align}
  \mb{Y}(l) = \mb{A}^{\tTrue}  \mb{H}^{\tTrue}(l) \mb{B}^{\tTrue\tT} \mb{X}(l) + \mb{W}(l) ,
  \label{eq:fdSignalSmv}
\end{align}
where $[\mb{Y}(l)]_{m,n}$ denotes the signal received by antenna $m$ on subcarrier $n$, for $m=1,\ldots,M$ and $n=1,\ldots,N$ (cmp. \cite{sparseMultipath}-\cite{sparseUnderwater}). In \eqref{eq:fdSignalSmv}, the $M \times P$ array steering matrix is given as
\begin{align}
  \mb{A}^{\tTrue} = [ \mb{a}(\theta^{\tTrue}_{1}), \ldots, \mb{a}(\theta^{\tTrue}_{P}) ],
\end{align}
where $\mb{a}(\theta) \!=\! [ 1, \e{-\j r_2 \xi(\theta) }, \scalebox{0.7}[1]{\ldots}, \e{-\j r_M \xi(\theta) } ]^\tT / \sqrt{M}$ denotes the normalized array steering vector for AoA $\theta$, with electric angle $\xi(\theta)=\frac{2\pi}{\lambda} \cos \theta$ and signal carrier wavelength $\lambda$. The $N \times P$ frequency response matrix is defined as
\begin{align}
  \mb{B}^{\tTrue} = [ \mb{b}(\tau^{\tTrue}_{1}), \ldots, \mb{b}(\tau^{\tTrue}_{P}) ]
\end{align}
with 
$\mb{b}(\tau) = [ 1, \e{-\j \frac{2 \pi}{N T_s} \tau  }, \scalebox{0.7}[1]{\ldots}, \e{-\j \frac{2 \pi (N-1)}{N T_s} \tau  } ]^\tT / \sqrt{N}$ denoting the normalized frequency response vector corresponding to a path delay of~$\tau$. The $P \times P$ diagonal channel gain matrix
\begin{align}
  \mb{H}^{\tTrue}(l) = \diag{h_{1}^{\tTrue}(l), \cdots, h_{P}^{\tTrue}(l)} 
\end{align}
contains the complex channel gain coefficients on its main diagonal and the $N \times N$ diagonal reference signal matrix
\begin{align}
  \mb{X}(l) = \diag{x_1(l), \ldots, x_N(l)}
\end{align}
contains the elements of the OFDM reference signal on its main diagonal. The elements of the $M \times N$ additive noise matrix $\mb{W}(l)$ represent spatially and temporally white Gaussian noise with variance $\sigma_w^2$. 

We collect $L$ snapshots, as defined in \eqref{eq:fdSignalSmv}, where we assume that the propagation delays $\{\tau_p^{\tTrue}\}_{p=1}^P$ and AoAs $\{\theta_p^{\tTrue}\}_{p=1}^P$ remain constant, while the channel gain coefficients $\{h_p^{\tTrue}(l)\}_{p=1}^P$ may vary between snapshots. Upon defining
\begin{align}
  \tilde{\mb{Y}} \!\!=\,& [\vec{\mb{Y}(1) \mb{X}^{-1}(1)}, \ldots, \vec{\mb{Y}(L)\mb{X}^{-1}(L)}], \label{eq:mmv1}\\
  \tilde{\mb{H}}\vphantom{\mb{H}}\vphantom{\mb{H}}^{\tTrue} \!\!=\,& [ \vecd{\mb{H}^{\tTrue}(1)}, \ldots, \vecd{\mb{H}^{\tTrue}(L)} ], \\
  \tilde{\mb{W}} \!\!=\,& [\vec{\mb{W}(1)\mb{X}^{-1}(1)}, \ldots, \vec{\mb{W}(L)\mb{X}^{-1}(L)}], \label{eq:mmv3}
\end{align}
where $\vec{\mb{Y}}$ denotes vectorization of matrix $\mb{Y}$ and $\vecd{\mb{H}}$ denotes vectorization of the elements on the main diagonal of $\mb{H}$, the multiple snapshot signal model is given by
\begin{align}
  \tilde{\mb{Y}} = ( \mb{B}^{\tTrue} \khatri \mb{A}^{\tTrue} ) \, \tilde{\mb{H}}\vphantom{\mb{H}}^{\tTrue}  + \tilde{\mb{W}}  
  \label{eq:fdSignalMmv}
\end{align}
where $\khatri$ denotes the Khatri-Rao product, i.e., the columnwise Kronecker product. 

\section{A Novel MD Sparse Recovery Method}
Following the ideas in \cite{mdRare}, we start by rewriting the signal model in \eqref{eq:fdSignalMmv} as
\begin{align}
  \tilde{\mb{Y}} =& ( \mb{B}^{\tTrue} \khatri \mb{A}^{\tTrue} ) \, \tilde{\mb{H}}\vphantom{\mb{H}}^{\tTrue}  + \tilde{\mb{W}} \nonumber \\
		 =& ( \mb{B}^{\tTrue} \otimes \mb{I}_M ) (\mb{I}_P \khatri \mb{A}^{\tTrue}) \, \tilde{\mb{H}}\vphantom{\mb{H}}^{\tTrue}  + \tilde{\mb{W}} \nonumber \\
		 =& ( \mb{B}^{\tTrue} \otimes \mb{I}_M ) \, \mb{G}_A^{\tTrue}  + \tilde{\mb{W}}
		 \label{eq:nucModel}
\end{align}
with $\otimes$ denoting the Kronecker product and $\mb{I}_M$ denoting the $M \times M$ identity matrix. The extended channel gain matrix 
\begin{align}
  \mb{G}_A^{\tTrue} = (\mb{I}_P \khatri \mb{A}^{\tTrue}) \tilde{\mb{H}}\vphantom{\mb{H}}^{\tTrue}
\end{align}
of size $MP \times L$ is composed of $P$ submatrices, i.e., $\mb{G}_A^{\tTrue} = [\mb{G}_{A,1}^{\tTrue\tT}, \ldots, \mb{G}_{A,P}^{\tTrue\tT}]^\tT$, where each $M \times L$ submatrix
\begin{align}
  \mb{G}_{A,p}^{\tTrue} = \mb{a}(\theta_p^{\tTrue}) \tilde{\mb{h}}\vphantom{\mb{h}}_p^{\tTrue\tT}
  \label{eq:doubleSparse1}
\end{align}
is of rank-one, with $\tilde{\mb{h}}\vphantom{\mb{h}}_p^{\tTrue}$ representing the $p$th row of $\tilde{\mb{H}}\vphantom{\mb{H}}^{\tTrue}$. 

For a sparse representation of \eqref{eq:nucModel} we uniformly discretize the delay parameter space in $Q$ points $\{ \tau_q \}_{q=1}^Q$, with $\max_q \tau_q < T_{\text{cp}}$, and form the $MN \times MQ$ dictionary matrix 
\begin{align}
  \mb{C}_B = [\mb{C}_{B,1}, \ldots, \mb{C}_{B,Q}] = \mb{B} \otimes \mb{I}_M,
\end{align}
with $\mb{B} = \left[ \mb{b}(\tau_1), \ldots, \mb{b}(\tau_Q) \right]$ representing the $N \times Q$ frequency response dictionary matrix, such that each submatrix $\mb{C}_{B,q} = \mb{b}(\tau_q) \otimes \mb{I}_M$, for $q=1,\ldots,Q$, contains the elements of the frequency response for path delay $\tau_q$. In correspondence with equations \eqref{eq:nucModel}-\eqref{eq:doubleSparse1} and under the assumption that the true propagation delays lie on the parameter grid, i.e., $\{ \tau_p^{\tTrue} \}_{p=1}^P~\in~\{ \tau_q \}_{q=1}^Q$, we formulate the block sparse matrix
\begin{align}
  \mb{G}_A = [\mb{G}_{A,1}^{\tT}, \ldots, \mb{G}_{A,Q}^{\tT}]^\tT
\end{align}
of size $MQ \times L$, with 
\begin{align}
  \mb{G}_{A,q} =
  \begin{cases}
	\mb{G}_{A,p}^{\tTrue} \quad & \text{ if } \tau_q = \tau_p^{\tTrue} \\
	\mb{0}			   & \text{ else},
  \end{cases}
  \label{eq:doubleSparse2}
\end{align}
such that we arrive at the sparse representation
\begin{align}
  \tilde{\mb{Y}} = \, \mb{C}_B \mb{G}_A  + \tilde{\mb{W}} .
		 \label{eq:nucModel2}
\end{align}
As can be seen from \eqref{eq:doubleSparse1} and \eqref{eq:doubleSparse2}, $\mb{G}_{A}$ exhibits two levels of sparsity, block sparsity and rank sparsity. As proposed in \cite{Steffens} the twofold sparse structure can be exploited in the following convex optimization problem
\begin{align}
  \hat{\mb{G}}_A = \arg \min_{\mb{G}_A} \; \frac{1}{2} \big\| \tilde{\mb{Y}} - \mb{C}_B \mb{G}_{A} \big\|_\tF^2 + 
				   \mu \sum_{q=1}^Q \big\| \mb{G}_{A,q} \big\|_{*}
  \label{eq:optProblem}
\end{align}
where $\mu > 0$ is a regularization parameter, determining the sparsity of the solution $\hat{\mb{G}}_A$, and the nuclear norm of the $q$th submatrix $\mb{G}_{A,q}$ can be computed as the sum of its singular values $\sigma_{q,i}$, with $i=1,\ldots,r$, and $r=\min (M,L)$, i.e.,
\begin{align}
  \left\| \mb{G}_{A,q} \right\|_{*} = {\textstyle \sum_{i=1}^{r}} \sigma_{q,i} .
\end{align}
While the minimization of the nuclear norm terms leads to rank sparsity in the form of low-rank submatrices $\mb{G}_{A,q}$, the nonlinear coupling in form of the nuclear norm terms leads to a block sparse structure, i.e., the elements in submatrices $\mb{G}_{A,q}$ are either jointly zero or jointly nonzero. 

The formulation in \eqref{eq:optProblem} can be considered as a one-dimensional parameter estimation problem, i.e. a block sparse representation in terms of the path delays. To recover all the channel coefficients from the matrix estimate $\hat{\mb{G}}_A$ in \eqref{eq:optProblem}, let us first define the support set as
\begin{align}
  \mathcal{I} = \big\{ q \big| \hat{\mb{G}}_{A,q} \neq \mb{0} \big\}, 
\end{align}
i.e., the indices of the nonzero submatrices $\hat{\mb{G}}_{A,q}$. Given the support set $\mathcal{I}$ of cardinality $\hat{P} = |\mathcal{I}|$, the estimated channel delays are found from the parameter grid $\{ \tau_q \}_{q=1}^Q$ as
\begin{align}
  \{ \hat{\tau}_p \}_{p=1}^{\hat{P}} = \left\{ \tau_q \big| q \in \mathcal{I} \right\} .
\end{align}
Furthermore, let the singular value decomposition of the $q$th nonzero submatrix be given as $\hat{\mb{G}}_{A,q} = \mb{U}_q \mb{\Sigma}_q \mb{V}_q^\tH$, then the corresponding AoAs $\hat{\theta}_q$ and complex channel gain coefficients $\hat{\tilde{\mb{h}}}_q$ can be estimated from the principal singular vectors: in the case of a rank-one submatrix $\hat{\mb{G}}_{A,q}$, the AoA estimation can be performed by 
\begin{align}
  \hat{\theta}_q = \textstyle \arg \max_{\theta} \; \big| \mb{a}^\tH(\theta) \mb{u}_{q,1} \big| \qquad \text{for } q \in \mathcal{I}
  \label{eq:estAoA}
\end{align}
where $\mb{u}_{q,1}$ is the principal left singular vector of $\hat{\mb{G}}_{A,q}$. In the case of higher rank submatrices, $r_q = \text{rank} (\hat{\mb{G}}_{A,q}) > 1$, there are $r_q$ propagation paths with path delays $\hat{\tau}_q = \ldots = \hat{\tau}_{q+r_q-1}$. The corresponding AoAs $\hat{\theta}_q, \ldots, \hat{\theta}_{q+r_q-1}$ are found from the $r_q$ array steering vectors $\mb{a}(\hat{\theta}_q), \ldots, \mb{a}(\hat{\theta}_{q+r_q-1})$ that best span the column subspace $\mb{U}_q$ of $\hat{\mb{G}}_{A,q}$ and can be estimated by standard methods, e.g., MUSIC or sparse recovery. Similarly, for a rank-one submatrix $\hat{\mb{G}}_{A,q}$, the complex channel gain coefficients in $\hat{\tilde{\mb{h}}\vphantom{\mb{H}}}_{q} = [\hat{h}_q(1), \ldots, \hat{h}_q(L)]^\tT$ can be recovered from the principal singular values and corresponding principal right singular vectors as
\begin{align}
  \hat{\tilde{\mb{h}}\vphantom{\mb{h}}}_{q} = \sigma_{q,1} \mb{v}^{*}_{q,1} / \sqrt{ML} \qquad \text{for } q \in \mathcal{I},
  \label{eq:channelGainEst}
\end{align}
such that all channel parameters of interest are paired with its corresponding propagation path. 

Note that alternatively the reformulation in \eqref{eq:nucModel} can be performed with respect to the array steering matrix $\mb{A}^{\tTrue}$, i.e.,
\begin{align}
  \tilde{\mb{Y}} =& ( \mb{B}^{\tTrue} \khatri \mb{A}^{\tTrue} ) \tilde{\mb{H}}\vphantom{\mb{H}}^{\tTrue}  + \tilde{\mb{W}} \nonumber \\
		 =& ( \mb{I}_N \otimes \mb{A}^{\tTrue} ) \mb{J} \mb{J}^\tT (\mb{B}^{\tTrue} \khatri \mb{I}_P ) \tilde{\mb{H}}\vphantom{\mb{H}}^{\tTrue}  + \tilde{\mb{W}} \nonumber \\
		 =& \mb{C}_A^{\tTrue} \mb{G}_B^{\tTrue}  + \tilde{\mb{W}}
\end{align}
where $\mb{C}_A^{\tTrue} = (\mb{I}_N \otimes \mb{A}^{\tTrue}) \mb{J}$ and $\mb{G}_B^{\tTrue} = (\mb{I}_P \khatri \mb{B}^{\tTrue}) \tilde{\mb{H}}\vphantom{\mb{H}}^{\tTrue}$, with proper permutation matrix $\mb{J}$, such that $\mb{J}^\tT (\mb{B}^{\tTrue}~\khatri~\mb{I}_P ) = \mb{I}_P~\khatri~\mb{B}^{\tTrue}$ and $\mb{J} \mb{J}^\tT = \mb{I}_{NP}$. Analogous to the steps \eqref{eq:nucModel}-\eqref{eq:channelGainEst} we can form a related estimation problem to recover the AoAs $\{ \hat{\theta}_p \}_{p=1}^{\hat{P}}$ in the first step and the delays $\{ \hat{\tau}_p \}_{p=1}^{\hat{P}}$ and complex channel gain coefficients $\{ \hat{\tilde{\mb{h}}}_p \}_{p=1}^{\hat{P}}$ in the second step. From our experiments we found that it is advantageous to perform the sparse representation with respect to the estimation parameter which provides more measurements, i.e., number of antennas and subcarriers. This brings benefits in estimation performance and reduces computational complexity, since the computation of the nuclear 
norm terms in
\eqref{eq:optProblem} is performed for smaller submatrices. 

Furthermore, we remark that similar to the signal models in \cite{1547647}-\nocite{mdRare}\cite{sparseUnderwater}, the model in \eqref{eq:nucModel} can easily be extended to incorporate more parameter dimensions, such as
two-dimensional angle-of-arrival, angle-of-departure or Doppler shift. The major advantage, that only discretization of one parameter space per successive estimation step is required for the estimation problem in \eqref{eq:optProblem} remains unchanged by this extension. 

\section{Algorithmic Implementation}
The nuclear norm minimization problem in \eqref{eq:optProblem} can be solved in various ways, e.g., by semidefintie programming \cite{Boyd:RankMinimization} or the block coordinate descent method \cite{Steffens}. One major challenge for the evaluation of the estimation problem \eqref{eq:optProblem} is the problem size, which cannot be solved with ease by the previously mentioned solvers. We propose to use the Soft-Thresholding with Exact Line search Algorithm (STELA) \cite{stela} since it is most suitable for solving large-scale sparse optimization problems.

For proper selection of the regularization parameter $\mu$ we further propose the computation of the solution path, i.e., the set of solutions to problem \eqref{eq:optProblem} for a sequence of regularization parameters $\{\mu_\kappa\}_{\kappa=1}^{\kappa_\text{max}}$, by means of the STELA method.

\subsection{STELA}
Here, we will shortly describe the major steps of STELA and refer to \cite{stela} for details. To simplify the notation we use $\mb{G} = \mb{G}_{A,B}$ and $\mb{C} = \mb{C}_{B,A}$. 

The key idea of STELA is to solve a sequence of approximate problems instead of the original problem in \eqref{eq:optProblem}. In iteration $t$, we approximate the problem in \eqref{eq:optProblem} by $Q$ independent subproblems of the following form
\begin{align}
  \mb{\Gamma}^{(t)}_q =& \arg \min_{\mb{\Gamma}_q} 
	\frac{1}{2} \big\| \tilde{\mb{Y}} - \mb{C}_{-q} \mb{G}^{(t)}_{-q} - \mb{C}_{q} \mb{\Gamma}_{q} \big\|_\tF^2 \nonumber \\
	& \phantom{\arg \min_{\mb{\Gamma}_q} } \; + \mu \textstyle{\sum_{q=1}^Q} ( \big\| \mb{G}^{(t)}_{-q} \big\|_{*} + \big\| \mb{\Gamma}_{q} \big\|_{*} ) ,
	\label{eq:coordOptProb}
\end{align}
for $q=1,\ldots,Q$, with 
$\mb{C}_{\!-q} = [ \mb{C}_{1}, \scalebox{0.9}[1]{\ldots}, \mb{C}_{q-1}, \mb{C}_{q+1}, \scalebox{0.9}[1]{\ldots}, \mb{C}_{Q} ]$ and 
$\mb{G}^{(t)}_{-q} = [ \mb{G}^{(t)\tT}_{1}, \scalebox{0.9}{\ldots}, \mb{G}^{(t)\tT}_{q-1}, \mb{G}^{(t)\tT}_{q+1}, \scalebox{0.9}[1]{\ldots}, \mb{G}^{(t)\tT}_{Q} ]^\tT$ 
denoting the dictionary matrix and an approximation of the optimal extended channel gain matrix in iteration $t$, with the $q$th submatrix removed, respectively. Note that in contrast to first or second order approximations of the objective function in \eqref{eq:optProblem}, the approximation in \eqref{eq:coordOptProb} relies on the concept of best-response, i.e., optimization for one submatrix while fixing the others. The matrix $\mb{\Gamma}^{(t)}_q$ in \eqref{eq:coordOptProb} denotes the $q$th submatrix of the best-response matrix $\mb{\Gamma}^{(t)} \!= \! [\mb{\Gamma}^{(t)\tT}_{1}\!, \scalebox{0.7}[1]{\ldots}, \mb{\Gamma}^{(t)\tT}_Q]^\tT$ in iteration $t$. 

For the case of unitary submatrices with $\mb{C}_{q}^\tH \mb{C}_{q} = \mb{I}$, as given in this work, it was shown in \cite{Steffens,Candes:SoftThresholding} that the problem \eqref{eq:coordOptProb} has the closed form solution
\begin{align}
  \mb{\Gamma}^{(t)}_q = \mathcal{S}_{\mu} ( \bar{\mb{\Gamma}}\vphantom{\mb{\Gamma}}^{(t)}_q ) 
  \label{eq:coordOptProb2}
\end{align}
where
\begin{align}
  \bar{\mb{\Gamma}}\vphantom{\mb{\Gamma}}^{(t)}_q = \mb{C}_q^\tH 
  ( \tilde{\mb{Y}} - \mb{C}_{-q} \mb{G}^{(t)}_{-q} )
\end{align}
is the Least-Squares estimate of $\mb{\Gamma}^{(t)}_q$, with compact singular value decomposition $\bar{\mb{\Gamma}}\vphantom{\mb{\Gamma}}^{(t)}_q = \bar{\mb{U}}^{(t)}_q \bar{\mb{\Omega}}_q^{(t)} \bar{\mb{V}}_q^{(t) \tH}$, and 
\begin{align}
  \mathcal{S}_{\mu} ( \mb{\Gamma}^{(t)}_q ) = 
  \bar{\mb{U}}^{(t)}_q \big( \bar{\mb{\Omega}}_q^{(t)} - \mu \mb{I} \big)_{+} \bar{\mb{V}}_q^{(t) \tH}
  \label{eq:minFun6}
\end{align}
denotes the singular value thresholding operator \cite{Candes:SoftThresholding}, with $\left[ ( \mb{X} )_{+} \right]_{ij} = \max (\left[ \mb{X} \right]_{ij}, 0)$. 

Returning to the original problem in \eqref{eq:optProblem}, it is shown in \cite{stela} that $\mb{\Gamma}^{(t)} - \mb{G}^{(t)}$ is a descent direction of the objective function in \eqref{eq:optProblem}. Therefore we perform a variable update according to
\begin{align}
  \mb{G}^{(t+1)} = \mb{G}^{(t)} + \gamma^{(t)} ( \mb{\Gamma}^{(t)} - \mb{G}^{(t)} ),
  \label{stelaUpdate}
\end{align}
with stepsize $\gamma^{(t)}$. Convergence of the sequence $\{\mb{G}^{(t)}\}_{t=1}^{\infty}$ to a stationary point strongly depends on proper selection of the stepsize parameter $\gamma^{(t)}$, e.g., by successive or exact line search methods. Following the ideas in \cite{stela}, exact line search is performed according to
\begin{align}
  \gamma^{(t)} \!=& \argmin_{0 \leq \gamma \leq 1} 
  \frac{1}{2} \big\| \tilde{\mb{Y}} - \mb{C} ( \mb{G}^{(t)} + \gamma (\mb{\Gamma}^{(t)} - \mb{G}^{(t)} )) \big\|_\tF^2 \nonumber \\
	&\phantom{\arg \min_{0 \leq \gamma \leq 1}} + 
	\gamma \mu \textstyle{\sum_{q=1}^Q} ( \| \mb{\Gamma}^{(t)}_q \|_{*} - \| \mb{G}^{(t)}_q \|_{*} ) \nonumber \\
  =& \Big[ \real\big\{\trace\big( ( \mb{C} \mb{G}^{(t)} \!- \tilde{\mb{Y}} )^\tH \mb{C} ( \mb{\Gamma}^{(t)} \!- \mb{G}^{(t)} ) \big)\big\}
	  \nonumber \\ & + 
	  \mu \textstyle{\sum_{q=1}^Q} ( \| \mb{\Gamma}^{(t)}_q \|_{*} \!- \| \mb{G}^{(t)}_q \|_{*} )
	  / \big\| \mb{C} ( \mb{\Gamma}^{(t)} \!\!- \mb{G}^{(t)} ) \big\|_\tF^2  \Big]_0^1 .
  \label{eq:stepSize}
\end{align}
In contrast to standard exact line search approaches, which generally have to be evaluated numerically, for problem \eqref{eq:optProblem} the stepsize parameter proposed in \eqref{eq:stepSize} can be computed in closed-form which significantly reduces the computational complexity. Similarly, as seen from eq. \eqref{eq:coordOptProb}-\eqref{eq:minFun6}, all submatrices $\mb{\Gamma}^{(t)}_q$, for $q=1,\ldots,Q$, admit closed form expressions and can be computed independently and, thus, in parallel. 

It can be verified that the problem in \eqref{eq:optProblem} and the approximate subproblems in \eqref{eq:coordOptProb} fulfill all assumptions specified by Theorem 2 in \cite{stela}, such that the sequence $\{\mb{G}^{(t)}\}_{t=1}^{\infty}$ converges to the global optimum $\hat{\mb{G}}$ of problem \eqref{eq:optProblem}. Furthermore, due to the exact line search, parallel updates and closed form expressions, the convergence speed of STELA is generally much faster than that of block coordinate descent methods or gradient-based methods, as we will show in the numerical results later.

In summary, one iteration of the STELA method consists of computing the submatrices of the best-response matrix $\mb{\Gamma}^{(t)}$ according to \eqref{eq:coordOptProb2}, the stepsize $\gamma\mysuper{(t)}$ according to \eqref{eq:stepSize} and updating the approximate solution $\mb{G}^{(t)}$ according to \eqref{stelaUpdate}. Initialization at $t=0$ can be performed by exploiting a-priori information or in the simplest form by $\mb{G}^{(0)}=\mb{0}$.

\subsection{Regularization Parameter Selection} \label{sec:regPar}
A key parameter determining the estimation performance of the minimization problem in \eqref{eq:optProblem} is the regularization parameter $\mu$. While direct estimation of a proper regularization parameter $\mu$ is still an open problem, in this work we follow the approach of computing the solution path to problem \eqref{eq:optProblem} as proposed in \cite{Steffens}. The solution path is defined as the set of solutions $\{\hat{\mb{G}}(\mu_\kappa)\}_{\kappa=1}^{\kappa_\text{max}}$ to \eqref{eq:optProblem} for a regularization parameter sequence $\mu_1 > \mu_2 > \ldots > \mu_{\kappa_\text{max}} >0$. 
Similar to the LASSO problem \cite{Osborne99onthe}, the starting point of the solution path can be computed in closed form as
\begin{align}
  \mu_1 = \textstyle \max_q \big\| \mb{C}_q^\tH \tilde{\mb{Y}} \big\|_2
\end{align}
which is the smallest regularization parameter giving a zero solution $\hat{\mb{G}}(\mu_1)=\mb{0}$ for problem \eqref{eq:optProblem}. 

A small change in $\mu_\kappa$ results in a small change in the estimate $\hat{\mb{G}}(\mu_{\kappa})$. This makes the STELA method particularly useful for the pathwise approach since the estimates $\hat{\mb{G}}(\mu_{\kappa-1})$ can be used as an initial value $\mb{G}^{(0)}(\mu_{\kappa})$ in the STELA iterations, leading to rapid convergence of the STELA method. 

Given the solution path $\{\hat{\mb{G}}(\mu_\kappa)\}_{\kappa=1}^{\kappa_\text{max}}$, we select the smallest regularization parameter $\mu_{\kappa}$ which generates a number of $P$ peaks in the spatial spectrum of the solution $\hat{\mb{G}}(\mu_\kappa)$. The corresponding estimate $\hat{\mb{G}}(\mu_{\kappa})$ represents $P$ dominant propagation paths with the best Least-Squares data fit in \eqref{eq:optProblem}. 

\section{Numerical Results}

We consider a uniform linear array of $M=4$ antennas. The OFDM reference signal consists of QPSK-symbols on $N=16$ subcarriers in the frequency domain and is sampled uniformly in time at the receiver. We assume that $L=3$ snapshots are available for parameter estimation. 

The complex channel gain coefficients are modeled as constant magnitude and uniform random phase. Let $(|h_{u,p}|, {\tau}_{p}/T_s, \theta_{p}/\text{deg.})$ denote the parameter triple defining the $p$th propagation path. We model the channel parameters as $(1.12, 0.10, 64.98)$, $(0.85, 1.23, 46.54)$, $(0.71, 1.97, 94.71)$, $(0.52, 3.57, 121.17)$ and $(0.41, 5.02, 105.32)$ such that there are $P=5$ propagation paths in total. Note that MD-ESPRIT is not directly applicable to this scenario, since it cannot resolve $P>\min(M,N)$ signals. The signal-to-noise ratio is defined as $\text{SNR} = 1/\sigma_w^2$. 

\pgfplotsset{minor grid style={densely dotted}}
\pgfplotsset{major grid style={densely dotted}}

\begin{figure}[b]
  \centering
  \footnotesize
  \vspace{-0.5cm}
  \input{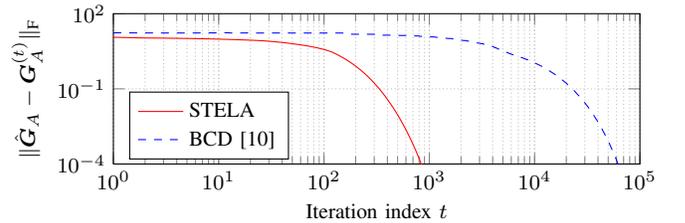}
  \vspace{-0.3cm}
  \caption{Convergence of STELA and block coordinate descent (BCD) method}
  \label{fig:convergence}
\end{figure}

In a first experiment we compare the convergence speed of STELA with that of the block coordinate descent (BCD) method \cite{Steffens}, by comparing the error $\|\hat{\mb{G}}_A - \mb{G}^{(t)}_A\|_\tF$, where $\hat{\mb{G}}_A$ denotes the solution to \eqref{eq:optProblem} and $\mb{G}^{(t)}_A$ denotes the approximate solution in iteration $t$. The BCD performs sequential submatrix updates, such that we count one submatrix update as an iteration, while for STELA we consider one parallel update of all submatrices as one iteration. For the simulation we assume an SNR of 5dB and select the regularization parameter as $\mu = \mu_1/4$ and a uniform grid of $Q=160$ points. As can be seen in Fig.~\ref{fig:convergence}, STELA clearly outperforms BCD in terms of convergence speed by an order of 2.

For statistical evaluation we compare the proposed method to the MD MUSIC estimator \cite{1547647}, the MD root-RARE estimator \cite{mdRare}, the space-alternating orthogonal matching pursuit (SA-OMP) \cite{sparseUnderwater} and the lower Cram\'{e}r-Rao bound (CRB) \cite{Zoltowski:Crb}. 

In order to achieve a sample covariance matrix of sufficiently high rank for the MUSIC and root-RARE estimator we employ  smoothing and forward-backward averaging in the frequency domain. We emphasize that these data pre-processing techniques require linear sampling, whereas our proposed method can be applied to arbitrary sampling schemes. 

For the proposed method we discretized the delay parameter space in $Q=160$ points and compute the solution path of \eqref{eq:optProblem} by means of STELA to find a proper estimate according to section \ref{sec:regPar}. Note that with the proposed  approach of regularization parameter selection, all evaluated methods exploit the same a-priori information in form of the number of propagation paths. To achieve higher resolution we perform a final estimation with a refined grid. The AoAs are estimated from the principal singular vectors according to equation \eqref{eq:estAoA}. 

\begin{figure}[t]
  \centering
  \footnotesize
%
%
\definecolor{mycolor1}{rgb}{1,0,1}%
\begin{tikzpicture}

\begin{axis}[%
width=7cm,
height=3.9cm,
scale only axis,
xmin=-10,
xmax=20,
xlabel={SNR in dB},
xmajorgrids,
ymode=log,
ymin=0.001,
ymax=10,
yminorticks=true,
ylabel={RMSE($\tau$)},
ymajorgrids,
yminorgrids,
legend style={draw=black,fill=white,legend cell align=left,row sep=-2pt,inner ysep=0pt},
xlabel near ticks,
ylabel near ticks
]
\addplot [
color=red,
solid,
mark=triangle,
mark options={solid}
]
table[row sep=crcr]{
-10 4.04800136088443\\
-5 1.79495403076582\\
0 0.323492422867621\\
5 0.0348448521922328\\
10 0.0189749570010478\\
15 0.0127126808108211\\
20 0.00872649445635872\\
};
\addlegendentry{Proposed};

\addplot [
color=green,
solid,
mark=asterisk,
mark options={solid}
]
table[row sep=crcr]{
-10 3.90154630324258\\
-5 2.03923864342539\\
0 0.654589048910047\\
5 0.0374883350766075\\
10 0.0216285496893342\\
15 0.0121645833888382\\
20 0.00673865752802437\\
};
\addlegendentry{MUSIC \cite{1547647}};

\addplot [
color=blue,
solid,
mark=o,
mark options={solid}
]
table[row sep=crcr]{
-10 5.31301944641107\\
-5 2.63775729601069\\
0 0.83478091574863\\
5 0.0997395139049563\\
10 0.0214660868626349\\
15 0.0121464129489322\\
20 0.00671899396726925\\
};
\addlegendentry{Root-RARE \cite{mdRare}};

\addplot [
color=orange,
solid,
mark=square,
mark options={solid}
]
table[row sep=crcr]{
-10 3.84202635905084\\
-5 1.41334740834161\\
0 0.0648992468908147\\
5 0.0468793212571957\\
10 0.0357853462601858\\
15 0.0370258803207462\\
20 0.0374863423167147\\
};
\addlegendentry{SA-OMP \cite{sparseUnderwater}};

\addplot [
color=black,
solid,
mark options={solid}
]
table[row sep=crcr]{
-10 0.192898797040442\\
-5 0.093120076443087\\
0 0.0493046973693598\\
5 0.0271575719817033\\
10 0.015169221611286\\
15 0.00851194661048767\\
20 0.00478335412146786\\
};
\addlegendentry{CRB \cite{Zoltowski:Crb}};

\end{axis}
\end{tikzpicture}%
  \vspace{-0.2cm}
  \caption{RMSE for the path delays}
  \label{fig:rmseDelay}
  \vspace{0cm}

%
%
\definecolor{mycolor1}{rgb}{1,0,1}%
\begin{tikzpicture}

\begin{axis}[%
width=7cm,
height=3.9cm,
scale only axis,
xmin=-10,
xmax=20,
xlabel={SNR in dB},
xmajorgrids,
ymode=log,
ymin=0.1,
ymax=100,
yminorticks=true,
ylabel={RMSE$(\theta)$},
ymajorgrids,
yminorgrids,
legend style={draw=black,fill=white,legend cell align=left,row sep=-2pt,inner ysep=0pt},
xlabel near ticks,
ylabel near ticks
]
\addplot [
color=red,
solid,
mark=triangle,
mark options={solid}
]
table[row sep=crcr]{
-10 44.9635707656765\\
-5 26.2017635742329\\
0 6.88940090283618\\
5 1.0176018867907\\
10 0.591342878540022\\
15 0.320417228001241\\
20 0.183141475368088\\
};
\addlegendentry{Proposed};

\addplot [
color=green,
solid,
mark=asterisk,
mark options={solid}
]
table[row sep=crcr]{
-10 26.4168550297925\\
-5 16.5334147087828\\
0 4.87108520127234\\
5 1.15767702818221\\
10 0.693562695707317\\
15 0.371572867487388\\
20 0.213234786865559\\
};
\addlegendentry{MUSIC \cite{1547647}};

\addplot [
color=blue,
solid,
mark=o,
mark options={solid}
]
table[row sep=crcr]{
-10 25.731297270544\\
-5 19.9159177869284\\
0 9.65135136102008\\
5 2.57888522693137\\
10 0.729091451324954\\
15 0.376935282127387\\
20 0.212327250435836\\
};
\addlegendentry{Root-RARE \cite{mdRare}};

\addplot [
color=orange,
solid,
mark=square,
mark options={solid}
]
table[row sep=crcr]{
-10 25.2635487804438\\
-5 12.7857119777882\\
0 1.88332180405693\\
5 1.18929654034645\\
10 0.919719791434794\\
15 0.770259097943575\\
20 0.721876552903284\\
};
\addlegendentry{SA-OMP \cite{sparseUnderwater}};

\addplot [
color=black,
solid,
mark options={solid}
]
table[row sep=crcr]{
-10 6.24688183025274\\
-5 3.02722701976534\\
0 1.60581498781742\\
5 0.885109852682466\\
10 0.494503490868055\\
15 0.277502617712591\\
20 0.155948407289054\\
};
\addlegendentry{CRB \cite{Zoltowski:Crb}};

\end{axis}
\end{tikzpicture}%
  \vspace{-0.2cm}
  \caption{RMSE for the path AoAs}
  \label{fig:rmseDoa}
  \vspace{-0.57cm}
\end{figure}

We perform 100 Monte Carlo runs and compute the root-mean-square errors (RMSEs) for the estimated propagation delays $\{\hat{\tau}_p\}_{p=1}^{\hat{P}}$ and AoAs $\{\hat{\theta}_p\}_{p=1}^{\hat{P}}$. As can be seen from Fig.~\ref{fig:rmseDelay} and \ref{fig:rmseDoa}, the SA-OMP scheme has the best thresholding performance among all evaluated methods. This is due to the joint estimation of delays and AoAs. However, for high SNR  the RMSE reaches an estimation bias, which can be explained by the greedy nature of the algorithm \cite{sparseUnderwater}. Similar to the SA-OMP method, the MUSIC method jointly estimates delays and AoAs. While the asymptotic behavior for high SNR approaches the CRB, the thresholding performance is slightly decreased as compared to SA-OMP, which is due to the low number of snapshots. Our proposed sparse estimation method has thresholding performance similar to the MUSIC estimator but has a much smaller complexity, due to the successive nature of the algorithm. Furthermore it 
outperforms the Root-RARE estimator in thresholding performance, which performs a similar successive parameter estimation. As can be seen from Fig.~\ref{fig:rmseDelay} and \ref{fig:rmseDoa} direct estimation of path delays and indirect estimation of the AoAs show equally good performance.

\section{Concluding Remarks}
We have presented a new method for MIMO channel parameter estimation based on nuclear norm minimization. In contrast to existing spectrum-based methods which perform large-scale and expensive multidimensional estimation, our proposed approach successively performs one-dimensional parameter estimation. The successive one-dimensional estimation requires discretization of only one parameter space at a time, which significantly reduces the computational complexity and the sensitivity to off-grid effects, while providing proper parameter pairing. Due to space limitations we have restricted the discussion to the two-dimensional parameter estimation problem, but the proposed approach can easily be extended to incorporate, e.g., two-dimensional angle-of-arrival, angle-of-departure or Doppler shift estimation (cmp. \cite{mdRare}). We conclude that the aforementioned benefits of our proposed method become even more significant in higher-dimensional parameter estimation problems. Furthermore, we have presented 
implementation based on the recently presented STELA method, which admits for fast convergence and simple implementation. 

\bibliographystyle{IEEEtran}
\bibliography{refferences}

\end{document}